\documentclass[aps,preprint]{revtex4}%
\usepackage{amsfonts}
\usepackage{subfigure}
\usepackage{amsmath}
\usepackage{amssymb}
\usepackage{graphicx}%
\begin{document}
\preprint{ }
\title[ ]{\textbf{Non-maximal Tripartite Entanglement Degradation of Dirac and Scalar
fields in Non-inertial frames}}
\author{Salman Khan$^{\dag}$}
\email{sksafi@comsats.edu.pk}
\author{Niaz Ali Khan$^{\ddag}$}
\author{M. K. Khan$^{\ddag}$}
\affiliation{$^{\dag}$Department of Physics, COMSATS Institute of Information Technology,
Chak Shahzad, Islamabad, Pakistan.}
\affiliation{$^{\ddag}$Department of Physics, Quaid-i-Azam University, Islamabad, Pakistan.}
\date{December 27, 2013}

\begin{abstract}
The $\pi$-tangle is used to study the behavior of entanglement of a nonmaximal
tripartite state of both Dirac and scalar fields in accelerated frame. For
Dirac fields, the degree of degradation with acceleration of both one-tangle
of accelerated observer and $\pi$-tangle, for the same initial entanglement,
is different by just interchanging the values of probability amplitudes. A
fraction of both one-tangles and the $\pi$-tangle always survives for any
choice of acceleration and the degree of initial entanglement. For scalar
field, the one-tangle of accelerated observer depends on the choice of values
of probability amplitudes and it vanishes in the range of infinite
acceleration, whereas for $\pi$-tangle this is not always true. The dependence
of $\pi$-tangle on probability amplitudes varies with acceleration. In the
lower range of acceleration, its behavior changes by switching between the
values of probability amplitudes and for larger values of acceleration this
dependence on probability amplitudes vanishes. Interestingly, unlike bipartite
entanglement, the degradation of $\pi$-tangle against acceleration in the case
of scalar fields is slower than for Dirac fields.



\end{abstract}
\maketitle

\section{Introduction}

One of the potential resources for all kinds of quantum information tasks is
entanglement. It is among the mostly investigated properties of many particles
systems. Since the beginning of the birth of the fields of quantum information
and quantum computation, it has been the pivot in different perspective to
bloom up these fields to be matured for technological purposes \cite{Many1}.
The recent development by mixing up the concepts of relativity theory with
quantum information theory brought to fore the relative behavior of
entanglement \cite{Alsing1,Alsing2,Alsing3,Fuentes}. These studies show that
entanglement not only depends on acceleration of the observer but also
strongly depends on statistics. For practical application in most general
scenario, it is essential to thoroughly investigate the behavior of
entanglement and hence of different protocols (such as teleportation) of
quantum information theory using different statistics in curved spacetime.

The observer dependent character of entanglement under various setup for
different kinds of fields have been studied by a number of authors. For
example, the entanglement between two modes of a free maximally entangled
bosonic and fermionic pairs is studied in \cite{Alsing2,Alsing3}, between to
modes of noninteracting massless scalar field is analyzed in \cite{Fuentes},
between free modes of a free scalar field is investigated in \cite{Adesso}.
Similarly, the dynamics of tripartite entanglement under different situation
for different fields has also been studied. For example, in Ref. \cite{Hwang}
the degradation of tripartite entanglement between the modes of free scalar
field due to acceleration of the observer is investigated. All these studies
are carried by taking single mode approximation. The behavior of entanglement
in accelerated frame beyond the single mode approximation is studied in Ref.
\cite{Bruchi}. The effect of decoherence on the behavior of entanglement in
accelerated frame is studied in Ref. \cite{Salman}. All these and many other
related works show that entanglement in the initial state is degraded when
observed from the frame of an accelerated observer.

On the other hand, there are studies which show, counter intuitively, that the
Unruh effect not only degrade entanglement shared between an inertial and an
accelerated observer but also amplify it. Ref. \cite{Montero} studies such
entanglement amplification for a particular family of states for scalar and
Grassman scalar fields beyond the single mode approximation. A similar
entanglement amplification is reported for fermionic system in Ref.
\cite{Kown}. There are a number of other good papers on the dynamics of
entanglement in accelerated frames which can be found in the list \cite{Pan2}.

It is well known that considering correlations between the modes of stationary
observer with both particle and anti-particle modes in the two causally
disconnected regions in the Rindler spacetime provides a broad view for
quantum communications tasks. Such considerations enable the stationary
observer to setup communication with either of the two disconnected regions or
with both at the same time \cite{Martin2}. This is possible by considering the
formalism of quantum communication in the limit of beyond single mode
approximation \cite{Bruchi}. In the same work it is shown that the single mode
approximation holds for some family of states under appropriate constraints.
On the other hand, it has also been suggested that the single mode
approximation is optimal for quantum communication between the stationary
observer and the accelerated observer \cite{Hosler}. For the purpose of this
paper we will use the later approach.

In this paper, we investigate the dependence of the behavior of a nonmaximal
tripartite entanglement of both Dirac and scalar fields on the acceleration of
the observer frame and on the entanglement parameter that describes the degree
of entanglement in the initial state. We show that the degradation of
entanglement with acceleration not only depends on the degree of initial
entanglement but also depends on the individual values of the normalizing
probability amplitudes of the initial state. We consider three observers
($i=A,B,C$), Alice Bob and Charlie, in Minkowski space such that each of them
observes only one part of the following nonmaximal initial tripartite
entangled state%
\begin{equation}
\left\vert \psi_{\omega_{A},\omega_{B},\omega_{C}}\right\rangle =\alpha
\left\vert 0_{\omega_{A}}\right\rangle _{A}\left\vert 0_{\omega_{B}%
}\right\rangle _{B}\left\vert 0_{\omega_{C}}\right\rangle _{C}+\sqrt
{1-\alpha^{2}}\left\vert 1_{\omega_{A}}\right\rangle _{A}\left\vert
1_{\omega_{B}}\right\rangle _{B}\left\vert 1_{\omega_{C}}\right\rangle _{C},
\label{1}%
\end{equation}
where $\left\vert m_{\omega_{i}}\right\rangle $ for $m\in(0,1)$ are the
Minkowski vacuum and first excited states with modes specified by the
subscript $\omega_{i}$ and $\alpha$ is a parameter that specify the degree of
entanglement in the initial state. Under the single mode approximation
\cite{Bruchi} $\omega_{A}\sim\omega_{B}\sim\omega_{C}=\omega,$ we can write
$\left\vert m_{\omega_{i}}\right\rangle =\left\vert m\right\rangle _{i}$.

Instead of being all the time in an inertial frame, if the frame of one of the
observers, say Charlie, suddenly gets some uniform acceleration $a$, then, the
Minkowski vacuum and excited states change from the perspective of the
accelerated observer. The appropriate coordinates for the viewpoint of an
accelerated observer are Rindler coordinates
\cite{AsPach,Martin,Bruchi,Brown2}. The Rindler spacetime for an accelerated
observer splits into two regions, $\mathrm{I}$ (right) and $\mathrm{II}$
(left), that are separated by Rindler horizon and thus are causally
disconnected from each other. The Rindler coordinates $(\tau,\xi)$ in region
$\mathrm{I}$ are defined in terms of the Minkowski coordinates $(t,x)$ as
follows%
\begin{equation}
t=\frac{1}{a}e^{a\xi}\sinh(a\tau),\qquad x=\frac{1}{a}e^{a\xi}\cosh(a\tau).
\end{equation}
An exact similar transformation holds between the coordinates for the Rindler
region $\mathrm{II}$, however, each coordinate differ by an overall minus
sign. These new coordinates allow us to perform a Bogoliubov transformation
between the Minkowski modes of a field and Rindler modes. The Rindler modes in
the two Rindler regions form a complete basis in terms of which the Minkowski
modes can be expanded. Thus any state in Minkowski space can be represented in
Rindler basis as well. However, an accelerated observer in Rindler region
$\mathrm{I}$ has no access to information in Rindler region $\mathrm{II}$. The
degree of entanglement of modes in each Rindler region with the modes of
inertial observers has its own dynamics. To study the behavior of entanglement
in one region, being inaccessible, the modes in other region becomes
irrelevant and thus need to be trace out.

The Minkowski annihilation operator of an arbitrary frequency, observed by
Alice, is related to the two Rindler regions' operators of frequency, observed
by Charlie, more directly through an intermediate set of modes called Unruh
modes \cite{Bruchi}. The Unruh modes analytically extend the Rindler region$I$
modes to region $II$ and the region$II$ modes to region $I$. Since the Unruh
modes exist over all Minkowski space, they share the same vacuum as the
Minkowski annihilation operators. An arbitrary Unruh mode for a give
acceleration is given by
\begin{equation}
C_{\omega}=q_{L}C_{\omega,L}+q_{R}C_{\omega,R},
\end{equation}
where $q_{L}$ and $q_{R}$ are complex numbers satisfying the relation
$\left\vert q_{L}\right\vert ^{2}+\left\vert q_{R}\right\vert ^{2}=1$ and the
appropriate relations for the left and right regions' operators are given by
\cite{Bruchi}
\begin{align}
C_{\omega,R} &  =\cosh r_{\omega}a_{\omega,I}-\sinh r_{\omega}a_{\omega
,II}^{\dag},\nonumber\\
C_{\omega,L} &  =\cosh r_{\omega}a_{\omega,II}-\sinh r_{\omega}a_{\omega
,I}^{\dag},
\end{align}
where $a$, $a^{\dag}$ are Rindler particle operators of scalar field in the
two regions. For Grassman case, the transformation relations are given by%
\begin{align}
C_{\omega,R} &  =\cos r_{\omega}c_{\omega,I}-\sinh r_{\omega}d_{\omega
,II}^{\dag},\nonumber\\
C_{\omega,L} &  =\cos r_{\omega}c_{\omega,II}-\sinh r_{\omega}d_{\omega
,I}^{\dag},
\end{align}
where $c$, $c^{\dag}$ and $d$, $d^{\dag}$ are respectively Rindler particle
and antiparticle operators. The dimensionless parameter $r_{\omega}$ appears
in these equations is discussed below. For the purpose of this paper, in order
to recover single mode approximation we will set $q_{R}=1$ and $q_{L}=0$.

From the viewpoint of accelerated observer, the Minkowski vacuum and excited
states of the Dirac field are found to be, respectively, given by
\cite{Alsing3}.%
\begin{equation}
\left\vert 0\right\rangle _{M}=\cos r\left\vert 0\right\rangle _{I}\left\vert
0\right\rangle _{II}+\sin r\left\vert 1\right\rangle _{I}\left\vert
1\right\rangle _{II}, \label{2}%
\end{equation}%
\begin{equation}
\left\vert 1\right\rangle _{M}=\left\vert 1\right\rangle _{I}\left\vert
0\right\rangle _{II}. \label{3}%
\end{equation}
Similarly, for scalar field the Minkowski vacuum and excited states are given
by%
\begin{equation}
\left\vert 0\right\rangle _{M}=\frac{1}{\cosh r}%
{\displaystyle\sum\limits_{n=0}^{\infty}}
\tanh^{n}r\left\vert n\right\rangle _{I}\left\vert n\right\rangle _{II},
\label{4}%
\end{equation}%
\begin{equation}
\left\vert 1\right\rangle _{M}=\frac{1}{\cosh^{2}r}%
{\displaystyle\sum\limits_{n=0}^{\infty}}
\sqrt{n+1}\tanh^{n}r\left\vert n+1\right\rangle _{I}\left\vert n\right\rangle
_{II}. \label{5}%
\end{equation}
In the above equations, $\left\vert \cdot\right\rangle _{I}$ and $\left\vert
\cdot\right\rangle _{II}$ are Rindler modes in the two causally disconnected
Rindler regions, $\left\vert n\right\rangle $ represents number states and $r$
is a dimensionless parameter that depends on acceleration of the moving
observer and modes frequency. For Dirac field, it is given by $\cos
r=(1+e^{-2\pi\omega c/a})^{-1/2}$ such that $0\leq r\leq\pi/4$ for $0\leq
a\leq\infty$ and for scalar field, it is defined as $\cosh r=(1-e^{-2\pi\omega
c/a})^{-1/2}$ such that $0\leq r\leq\infty$ for $0\leq a\leq\infty$. It is
important to note that almost all the previous studies have been focused on
investigating the influence of parameter $r$, as a function of acceleration of
the moving frame by fixing the Rindler frequency, on the degree of
entanglement present in the initial state. Such analysis lead to the
measurement of entanglement in a family of states, all of which share the same
Rindler frequency as seen by an observer with different acceleration. However,
the effect of parameter $r$ on entanglement can also, alternatively, be
interpreted by considering a family of states with different Rindler
frequencies watched by the same observer traveling with fixed acceleration
\cite{Bruschi2}.

\section{Quantification of Tripartite entanglement}

In literature, a number of different criterion for quantifying tripartite
entanglement exist. However, the most popular among them are the residual
three tangle \cite{Coffman} and $\pi$-tangle \cite{Fan,Vidal}. Other
measurements for tripartite entanglement include realignment criterion
\cite{Rudolph,Kia} and linear contraction \cite{Horodocki}. The realignment
and linear contraction criterion are comparatively easy in calculation and are
strong criteria for entanglement measurement. However, these criterion has
some limitations and do not detect the entanglement of all states.

The three tangle is another good quantifier for the entanglement of tripartite
states. This is polynomial invariant \cite{Verstraete, Leifer} and it needs an
optimal decomposition of a mixed density matrix. In general, the optimal
decomposition is a tough enough task except in a few rare cases
\cite{Lohmayer}. On the other hand, the $\pi$-tangle for a tripartite state
$|\psi\rangle_{ABC}$ is given by%
\begin{equation}
\pi_{ABC}=\frac{1}{3}(\pi_{A}+\pi_{B}+\pi_{C}), \label{6}%
\end{equation}
where $\pi_{A}$ is called residual entanglement and is given by%
\begin{equation}
\pi_{A}=\mathcal{N}_{A(BC)}^{2}-\mathcal{N}_{AB}^{2}-\mathcal{N}_{AC}^{2}.
\label{7}%
\end{equation}
The other two residual tangles ($\pi_{B},\pi_{C}$) are defined in a similar
way. In Eq. (\ref{7}), $\mathcal{N}_{AB}(\mathcal{N}_{AC})$ is a two-tangle
and is given as the negativity of mixed density matrix $\rho_{AB}=Tr_{C}%
|\psi\rangle_{ABC}\langle\psi|$ $(\rho_{AC}=Tr_{B}|\psi\rangle_{ABC}%
\langle\psi|)$. The $\mathcal{N}_{A(BC)}$ is a one-tangle and is defined as
$\mathcal{N}_{A(BC)}=\left\Vert \rho_{ABC}^{T_{A}}\right\Vert -1$, where
$\left\Vert O\right\Vert =\mathrm{tr}\sqrt{OO^{\dag}}$ stands for the trace
norm of an operator $O$ and $\rho_{ABC}^{T_{A}}$ is the partial transposition
of the density matrix over qubit $A$. The one-tangle and the two-tangles
satisfy the following Coffman-Kundu-Wootters (CKW) monogamously inequality
relation \cite{Coffman}.%
\begin{equation}
\mathcal{N}_{A(BC)}^{2}\geq\mathcal{N}_{AB}^{2}+\mathcal{N}_{AC}^{2}.
\label{8}%
\end{equation}
In this paper we use the $\pi$-tangle to observe the behavior of entanglement
of the state given in Eq. (\ref{1}), as a function of acceleration of the
observer and the entanglement parameter $\alpha$.

\section{Nonmaximal tripartite entanglement}

\subsection{Fermionic Entanglement}

To study the influence of acceleration parameter $r$ and the entanglement
parameter $\alpha$ on the entanglement between modes of Dirac field, we
substitute Eqs.(\ref{2}) and (\ref{3}) for Charlie part in Eq.(\ref{1}) and
rewrite it in terms of Minkowski modes for Alice and Bob and Rindler modes for
Charlie as follow%
\begin{equation}
\left\vert \psi_{ABCI,II}\right\rangle =\alpha(\cos r\text{ }\left\vert
0000\right\rangle +\sin r\text{ }\left\vert 0011\right\rangle )+\sqrt
{1-\alpha^{2}}\left\vert 1110\right\rangle , \label{9}%
\end{equation}
where $\left\vert abcd\right\rangle =\left\vert a\right\rangle _{A}\left\vert
b\right\rangle _{B}\left\vert c\right\rangle _{CI}\left\vert d\right\rangle
_{CII}$. Note that for the purpose of writing ease, we have also dropped the
frequency in the subscript of each ket. Being inaccessible to Charlie in
Rindler region \textrm{I}, the modes in Rindler region \textrm{II} must be
trace out for investigating the behavior of entanglement between the modes of
inertial observers and the modes of Charlie in region \textrm{I}. So, tracing
out over the forth qubit, leaves the following mixed density matrix between
the modes of Alice, Bob and Charlie,%
\begin{align}
\rho_{ABC}  &  =\alpha^{2}\cos^{2}r\left\vert 000\right\rangle \left\langle
000\right\vert +\alpha\sqrt{1-\alpha^{2}}\cos r(\left\vert 000\right\rangle
\left\langle 111\right\vert +\left\vert 111\right\rangle \left\langle
000\right\vert )\nonumber\\
&  +\alpha^{2}\sin^{2}r\left\vert 001\right\rangle \left\langle 001\right\vert
+(1-\alpha^{2})\left\vert 111\right\rangle \left\langle 111\right\vert .
\label{10}%
\end{align}
Taking partial transpose over each qubit in sequence and using the definition
of one-tangle, the three one-tangles can straightforwardly be calculated,
which are given by%
\begin{equation}
\mathcal{N}_{A(BC)}=\mathcal{N}_{B(AC)}=2\alpha\sqrt{1-\alpha^{2}}\cos r.
\label{11}%
\end{equation}%
\begin{equation}
\mathcal{N}_{C(AB)}=\alpha\sqrt{1-\alpha^{2}}\cos r-\alpha^{2}\sin^{2}%
r+\alpha\sqrt{(1-\alpha^{2})\cos^{2}r+\alpha^{2}\sin^{4}r}. \label{12}%
\end{equation}
Note that $\mathcal{N}_{A(BC)}=\mathcal{N}_{B(AC)}$ shows that the two
subsystems of inertial frames are symmetrical for any values of the parameters
$\alpha$ and $r$. It can easily be checked that all the one-tangles reduce to
$1$ for a maximally entangled initial state with no acceleration, which is a
verification of the result obtained in the rest frames both for Dirac and
Scalar fields \cite{Hwang, Wang}. To have a better understanding of the
influence of the two parameters, we plot the one-tangles for different values
of $\alpha$ against $r$ in Fig. $1$(a, b).\begin{figure}[h]
\begin{center}
\subfigure[]{
\includegraphics[scale=0.9]{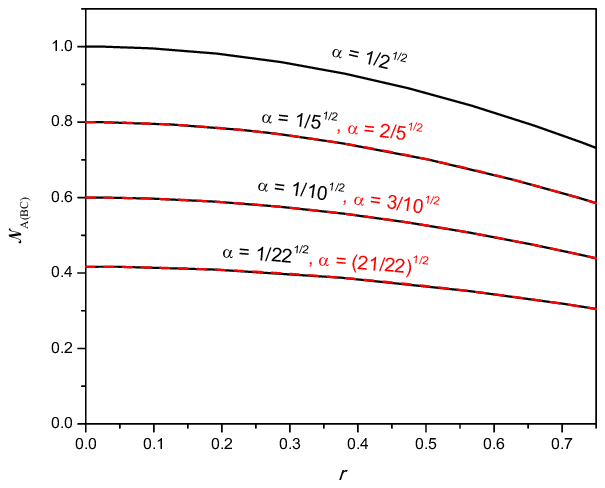}} \subfigure[]{
\includegraphics[scale=0.9]{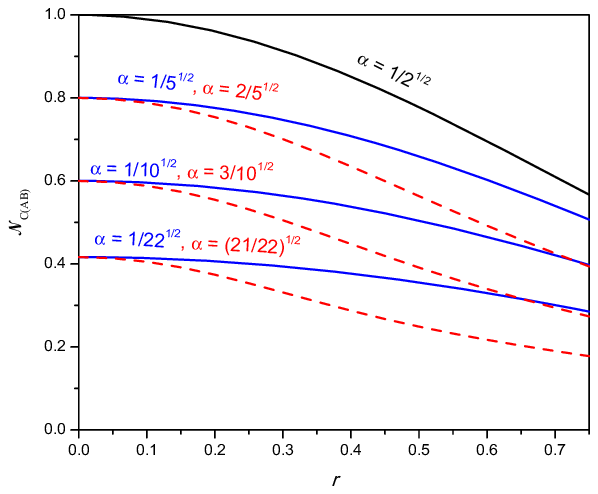}}
\end{center}
\caption{(Color Online) The one-tangles (a) $\mathcal{N}_{A(BC)}$ and (b)
$\mathcal{N}_{C(AB)}$ of fermionic modes as a function of the acceleration
parameter $r$\ for different values of the entanglement parameter $\alpha
$\ and its normalized partners\ $\sqrt{1-\alpha^{2}}$. The black solid line
corresponds to maximally entangled initial state. The blue solid lines from
top to bottom correspond to $\left\vert \alpha\right\vert =\frac{1}{\sqrt{5}}%
$, $\frac{1}{\sqrt{10}}$, $\frac{1}{\sqrt{22}}$ and red dashed lines from top
to bottom correspond to $\left\vert \alpha\right\vert =\frac{2}{\sqrt{5}}$,
$\frac{3}{\sqrt{10}}$, $\sqrt{\frac{21}{22}}$.}%
\label{Figure1}%
\end{figure}Figure (1a) shows the behavior of $\mathcal{N}_{A(BC)}%
=\mathcal{N}_{B(AC)}$ and figure (1b) is the plot of $\mathcal{N}_{C(AB)}$. A
comparison of the two figures shows that for maximal entangled initial state
($\alpha=1/\sqrt{2}$) and hence for all other values of $\alpha$, the
$\mathcal{N}_{C(AB)}$ falls off rapidly with increasing acceleration as
compared to $\mathcal{N}_{A(BC)}$. However, the most interesting feature of
the two figures is the different response of the one-tangles to the parameter
$\alpha$. The behavior of $\mathcal{N}_{A(BC)}$ ($\mathcal{N}_{B(AC)}$) is
unchanged by interchanging the values of $\alpha$ and its normalizing partner
$\sqrt{1-\alpha^{2}}$. On the other hand, $\mathcal{N}_{C(AB)}$\ degrades
along different trajectories by switching the values of $\alpha$ and
$\sqrt{1-\alpha^{2}}$. This shows an inequivalence of the quantization for
Dirac field in the Minkowski and Rindler coordinates. Regardless of the amount
of acceleration, there is always some amount of one-tangle left for each
subsystem, which ensures the application of entanglement based quantum
information tasks between relatively accelerated parties. The values chosen
for entanglement parameter $\alpha$ and its normalizing partner $\sqrt
{1-\alpha^{2}}$ in figure (1) are $\frac{1}{\sqrt{2}},\frac{1}{\sqrt{5}}%
,\frac{2}{\sqrt{5}},\frac{1}{\sqrt{10}},\frac{3}{\sqrt{10}},\frac{1}{\sqrt
{22}},\sqrt{\frac{21}{22}}$.

The next step is to evaluate the two-tangles. According to its definition, we
need to take partial trace over each qubit one by one. So, taking partial
trace of the final density matrix of Eq. (\ref{10}) over Alice's qubit or
Bob's qubit leads to the following mixed density matrix%
\begin{equation}
\rho_{AC(BC)}=\rho_{ABC}^{T_{B(A)}}=\alpha^{2}\cos^{2}r\left\vert
00\right\rangle \left\langle 00\right\vert +\alpha^{2}\sin^{2}r\left\vert
01\right\rangle \left\langle 01\right\vert +(1-\alpha^{2})\left\vert
11\right\rangle \left\langle 11\right\vert . \label{13}%
\end{equation}
Note that this matrix is diagonal and the partial transpose over either qubit
leaves it unchanged. Similarly, the reduced density matrix $\rho_{AB}$, which
is obtained by taking partial trace over the Charlie qubit, is diagonal. Using
the definition of negativity, one can easily show that there exists no
entanglement between any of these subsystems of the tripartite state
$\rho_{ABC}$. Since this result is valid for a maximally entangled GHZ state
in inertial frame, it shows that the entanglement behavior of subsystems is
independent from the status of the observer and from the degree of initial
entanglement in the state. Also, the zero value of all the two-tangles verify
the validity of the CKW inequality.

Since we now know all the one-tangles and all the two-tangles of the
tripartite state $\rho_{ABC}$, we can find the $\pi$-tangle. As all the
two-tangles are zero, using Eq.\ (\ref{6}), it simply becomes%
\begin{align}
\pi_{ABC}  &  =\frac{1}{3}(\mathcal{N}_{A(BC)}^{2}+\mathcal{N}_{B(AC)}%
^{2}+\mathcal{N}_{C(AB)}^{2})\nonumber\\
&  =\frac{\alpha^{2}}{3}[\left(  \sqrt{(1-\alpha^{2})}\cos^{2}r-\alpha\sin
^{2}r+\sqrt{(1-\alpha^{2})\cos^{2}r+\alpha^{2}\sin^{4}r}\right)
^{2}\nonumber\\
&  +8(1-\alpha^{2})\cos^{2}r]. \label{14}%
\end{align}
\begin{figure}[h]
\begin{center}
\includegraphics[scale=1.2]{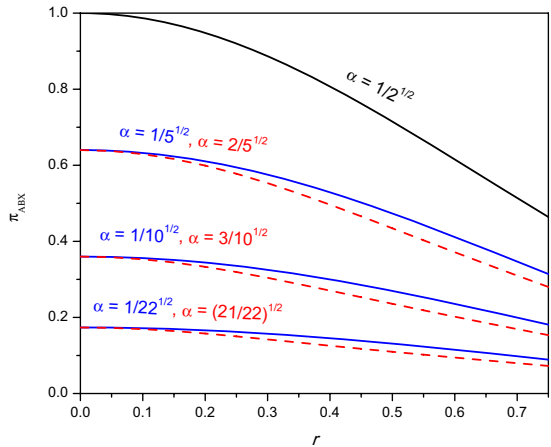}
\end{center}
\caption{(Color Online) The $\pi$-tangle\ of fermionic modes as a function of
acceleration parameter $r$\ for different values of entanglement parameter
$\alpha$\ and its normalized partner\ $\sqrt{1-\alpha^{2}}$. The black solid
line corresponds to maximally entangled initial state. The blue solid lines
from top to bottom correspond to $\left\vert \alpha\right\vert =\frac{1}%
{\sqrt{5}}$, $\frac{1}{\sqrt{10}}$, $\frac{1}{\sqrt{22}}$ and the red dashed
lines from top to bottom correspond to $\left\vert \alpha\right\vert =\frac
{2}{\sqrt{5}}$, $\frac{3}{\sqrt{10}}$, $\sqrt{\frac{21}{22}}$.}%
\label{Figure2}%
\end{figure}It is straightforward to verify that for inertial frame and
maximally entangled initial state the result of Eq. (\ref{14}) is $1$. To have
a more close look on how it is effected by the parameters $\alpha$ and $r$, we
plot it against the parameter $r$ for different values of the entanglement
parameter $\alpha$ in Fig. $2$. Like the one-tangles, the $\pi$\textit{-}%
tangle exhibit a similar behavior in response to $\alpha$. Here the solid
black line represents the behavior of $\pi$-tangle against $r$ when the
initial state is maximally entangled. It can be seen that for the same
entanglement in the initial state, interchanging the values of $\alpha$ and
its normalizing partner $\sqrt{1-\alpha^{2}}$ leads to two different
degradation curves for $\pi$-tangle against the acceleration parameter $r$.
This degradation behavior of $\pi$-tangle along two different curves is
similar to the degradation of logarithmic negativity for bipartite fermionic
entangled states \cite{Pan}.\ It is interesting to note that the loss of
entanglement against the acceleration parameter is rapid for states of
stronger initial entanglement. Nevertheless, the rate of degradation of $\pi
$-tangle is slower than the logarithmic negativity for bipartite fermionic states.

\subsection{Bosonic Entanglement}

To study the behavior of entanglement of nonmaximal initial state of scalar
field, we follow the same procedure as we used to investigate the dynamics of
entanglement of Dirac Field. For Charlie in noninertial frame, the nonmaximal
entangled initial state of Eq. (\ref{1}) can be rewritten in terms of
Minkowski modes for Alice and Bob and Rindler modes of Fock space for Charlie
by using Eqs. (\ref{4}) and (\ref{5}) as follow%
\begin{equation}
\left\vert \varphi_{ABCI,II}\right\rangle =\frac{1}{\cosh r}%
{\displaystyle\sum\limits_{n=0}^{\infty}}
\tanh^{n}r\left[  \alpha\left\vert 00nn\right\rangle +\frac{\sqrt
{(n+1)(1-\alpha^{2})}}{\cosh r}\left\vert 11n+1n\right\rangle \right]  ,
\label{15}%
\end{equation}
where, again, the kets $\left\vert abcd\right\rangle =\left\vert
a\right\rangle _{A}\left\vert b\right\rangle _{B}\left\vert c\right\rangle
_{CI}\left\vert d\right\rangle _{CII}$. In response to acceleration, for the
behavior of entanglement between the modes of inertial observers and the modes
of Charlie in region \textrm{I}, the inaccessible modes in region \textrm{II}
must be trace out. Tracing out over those modes, leaves the following mixed
density matrix%
\begin{align}
\varrho_{ABC}  &  =\alpha^{2}\left\vert 00\right\rangle \left\langle
00\right\vert \otimes M_{n,n}+(1-\alpha^{2})\left\vert 11\right\rangle
\left\langle 11\right\vert \otimes M_{n+1,n+1}+\nonumber\\
&  \alpha\sqrt{(1-\alpha^{2})}(\left\vert 11\right\rangle \left\langle
00\right\vert \otimes M_{n+1,n}+\left\vert 00\right\rangle \left\langle
11\right\vert \otimes M_{n,n+1}), \label{16}%
\end{align}
where%
\begin{align}
M_{n,n}  &  =\frac{1}{\cosh^{2}r}%
{\displaystyle\sum\limits_{n=0}^{\infty}}
\tanh^{2n}r\left\vert n\right\rangle \left\langle n\right\vert ,\nonumber\\
M_{n,n+1}  &  =\frac{1}{\cosh^{3}r}%
{\displaystyle\sum\limits_{n=0}^{\infty}}
\sqrt{(n+1)}\tanh^{2n}r\left\vert n\right\rangle \left\langle n+1\right\vert
,\nonumber\\
M_{n+1,n}  &  =\frac{1}{\cosh^{3}r}%
{\displaystyle\sum\limits_{n=0}^{\infty}}
\sqrt{(n+1)}\tanh^{2n}r\left\vert n+1\right\rangle \left\langle n\right\vert
,\nonumber\\
M_{n+1,n+1}  &  =\frac{1}{\cosh^{4}r}%
{\displaystyle\sum\limits_{n=0}^{\infty}}
(n+1)\tanh^{2n}r\left\vert n+1\right\rangle \left\langle n+1\right\vert .
\label{17}%
\end{align}
The three one-tangles can be computed, as before, by taking partial transpose
of the density matrix of Eq. (\ref{16}) with respect to each qubit one by one.
It is easy to prove that the two one-tangles which are obtained from partial
transposed of the qubits of inertial observers are equal and is given by%
\begin{equation}
\mathcal{N}_{A(BC)}=\mathcal{N}_{B(AC)}=\frac{2\alpha\sqrt{1-\alpha^{2}}%
}{\cosh^{3}r}\sum_{n=0}^{\infty}\sqrt{(n+1)}\tanh^{2n}r. \label{18}%
\end{equation}
We can write this relation into another more compact form as follow%
\begin{equation}
\mathcal{N}_{A(BC)}=\frac{2\alpha\sqrt{1-\alpha^{2}}}{\cosh r\sinh^{2}%
r}\mathbf{Li}_{-\frac{1}{2}}(\tanh^{2}r), \label{19}%
\end{equation}
where we have used the following identities%
\begin{align}
\sum_{n=0}^{\infty}(n+1)\tanh^{2n}r  &  =\cosh^{4}r\nonumber\\
\sum_{n=0}^{\infty}\tanh^{2n}r  &  =\cosh^{2}r. \label{20}%
\end{align}
The function $\mathbf{Li}_{n}(x)$ in Eq. (\ref{19}) is a polylogarithm
function and is given by%
\begin{equation}
\mathbf{Li}_{n}(x)\equiv\sum_{k=1}^{\infty}\frac{x^{k}}{k^{n}}=\frac{x}{1^{n}%
}+\frac{x^{2}}{2^{n}}+\frac{x^{3}}{3^{n}}+... \label{21}%
\end{equation}
To compute the one tangle $\mathcal{N}_{C(AB)}$, first we find $\varrho
_{ABC}^{T_{C}}$ from Eq.(\ref{16}) and then we construct $(\varrho
_{ABC}^{T_{C}})(\varrho_{ABC}^{T_{C}})^{\dag}$, whose explicit expression is
given by%
\begin{align}
(\varrho_{ABC}^{T_{C}})(\varrho_{ABC}^{T_{C}})^{\dag}  &  =\sum_{n=0}^{\infty
}\frac{\tanh^{4n}r}{\cosh^{4}r}[(\alpha^{4}+\frac{n\alpha^{2}(1-\alpha
^{2})\cosh^{2}r}{\sinh^{4}r})\left\vert 00n\right\rangle \left\langle
00n\right\vert +\frac{\alpha((n+1)(1-\alpha^{2})x)^{\frac{1}{2}}}{\cosh
r}\nonumber\\
&  (\alpha^{2}\tanh^{2}r+\frac{n(1-\alpha^{2})}{\sinh^{2}r})\{\left\vert
00n+1\right\rangle \left\langle 11n\right\vert +\left\vert 11n\right\rangle
\left\langle 00n+1\right\vert \}\nonumber\\
&  +(\frac{\alpha^{2}(1-\alpha^{2})(n+1)}{\cosh^{2}r}+\frac{n^{2}(1-\alpha
^{2})^{2}}{\sinh^{4}r})\left\vert 11n\right\rangle \left\langle 11n\right\vert
]. \label{22}%
\end{align}
The nonvanishing eigenvalues Eq. (\ref{22}) are%
\begin{equation}
\left(  \frac{\alpha^{4}}{\cosh^{4}r},\Lambda_{n}^{\pm},\text{ \ \ }%
(n=0,1,2,3,...)\right)  , \label{23}%
\end{equation}
where%
\begin{equation}
\Lambda_{n}^{\pm}=\frac{1}{2}(\xi\pm\sqrt{\eta+\mu}), \label{24}%
\end{equation}
and
\begin{align}
\xi &  =\frac{\tanh^{4n}r}{\cosh^{4}r}\left(  \frac{n^{2}(1-\alpha^{2})^{2}%
}{\sinh^{4}r}+\frac{2\alpha^{2}(1-\alpha^{2})(n+1)}{\cosh^{2}r}+\alpha
^{4}\tanh^{4}r\right)  ,\nonumber\\
\mu &  =\frac{4\alpha^{2}(1-\alpha^{2})(n+1)}{\cosh^{2}r}\frac{\tanh^{8n}%
r}{\cosh^{8}r}\left(  \frac{n(1-\alpha^{2})}{\sinh^{2}r}+\alpha^{2}\tanh
^{2}r\right)  ^{2},\nonumber\\
\eta &  =\frac{\tanh^{8n}r}{\cosh^{8}r}\left(  \frac{n^{2}(1-\alpha^{2})^{2}%
}{\sinh^{4}r}-\alpha^{4}\tanh^{4}r\right)  ^{2}. \label{25}%
\end{align}
Using the definition of one-tangle, one can obtain $\mathcal{N}_{C(AB)}$ whose
explicit expression is by%
\begin{equation}
\mathcal{N}_{C(AB)}=-1+\frac{\alpha^{2}}{\cosh^{2}r}+\sum_{n=0}^{\infty}%
\frac{\tanh^{2n}r}{\cosh^{2}r}\sqrt{\frac{n^{2}(1-\alpha^{2})^{2}}{\sinh^{4}%
r}+\frac{2\alpha^{2}(1-\alpha^{2})(n+2)}{\cosh^{2}r}+\alpha^{4}\tanh^{4}r}
\label{26}%
\end{equation}
It is easy to check that the one-tangles results into $1$ for $r=0$ and
maximally entangled initial state.\begin{figure}[h]
\begin{center}
\subfigure[]{
\includegraphics[scale=0.9]{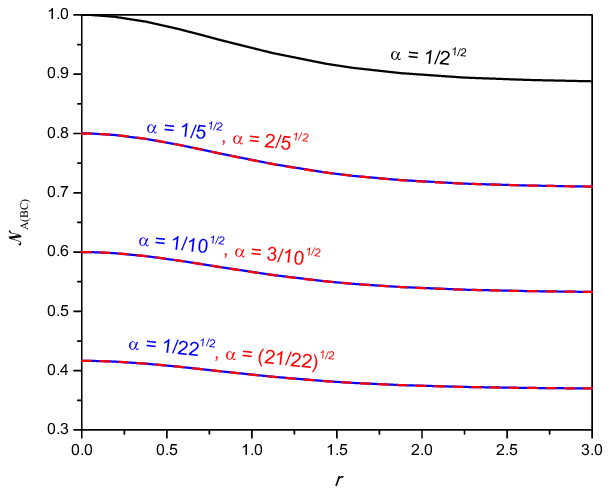}} \subfigure[]{
\includegraphics[scale=0.9]{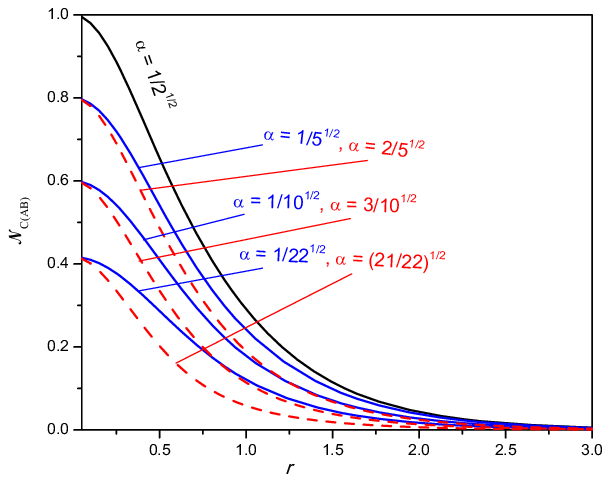}}
\end{center}
\caption{(Color Online)The one-tangle (a) $\mathcal{N}_{A(BC)}$ and (b)
$\mathcal{N}_{C(AB)}$ of bosonic field as a function of the acceleration
parameter $r$\ for different values of entanglement parameter $\alpha$\ and
its normalized partners\ $\sqrt{1-\alpha^{2}}$. The black solid line
corresponds to maximally entangled initial state. The blue solid lines from
top to bottom correspond to $\left\vert \alpha\right\vert =\frac{1}{\sqrt{5}}%
$, $\frac{1}{\sqrt{10}}$, $\frac{1}{\sqrt{22}}$ and the red dashed lines from
top to bottom correspond to $\left\vert \alpha\right\vert =\frac{2}{\sqrt{5}}%
$, $\frac{3}{\sqrt{10}}$, $\sqrt{\frac{21}{22}}$.}%
\label{Figure3}%
\end{figure}The dependence of one-tangles on $r$ and $\alpha$, in this case,
is shown in figure ($3$). As can be seen, the one-tangles are strongly
effected by the parameters $\alpha$ and $r$. However, as before, switching
between the values of $\alpha$ and its normalizing partner $\sqrt{1-\alpha
^{2}}$ does not effect the behavior of one-tangle, corresponds to an inertial
observer, against $r$ as shown in figure ($3a$). Unlike the fermionic case,
the loss in one-tangle $\mathcal{N}_{A(BC)}$ with acceleration is not uniform
through the whole range of $r$. In fermionic case, it is monotonic strictly
decreasing whereas in bosonic case, it is only monotonic decreasing, however,
it never vanishes completely. On the other hand, figure ($3b$) shows that,
like the fermionic case, the one-tangle $\mathcal{N}_{C(AB)}$ degrades along
different curves against $r$ by interchanging the values of $\alpha$ and
$\sqrt{1-\alpha^{2}}$, however, it vanishes , regardless of the value of
$\alpha$, in the asymptotic limit. The loss in $\mathcal{N}_{C(AB)}$ against
$r$ depends on the degree of entanglement in the initial state, it is faster
when the entanglement is stronger initially.

Similar to the case of Dirac field, we have verified that all the two tangles
for scalar field are also zero, that is,%
\begin{equation}
\mathcal{N}_{AB}=\mathcal{N}_{AC}=\mathcal{N}_{BC}=0. \label{27}%
\end{equation}
This verifies that CKW inequality also holds for scalar field. Again, the zero
values of all the two tangles make it easier to find the $\pi$-tangle. Instead
of writing its explicit relation, which is lengthy enough, we want to show its
behavior by plotting it against $r$ for different values of $\alpha$ in figure
($4$).\begin{figure}[h]
\begin{center}
\includegraphics[scale=1.2]{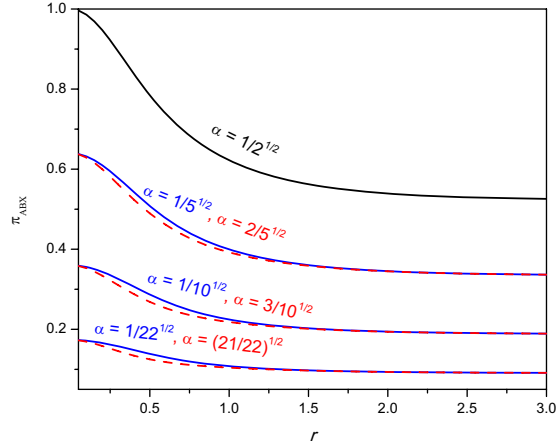}
\end{center}
\caption{(Color Online) The $\pi$-tangle\ of bosonic field as a function of
acceleration parameter $r$\ for different values of entanglement parameter
$\alpha$\ and its normalized partners\ $\sqrt{1-\alpha^{2}}$. The black solid
line corresponds to maximally entangled initial state. The blue solid lines
from top to bottom correspond to $\left\vert \alpha\right\vert =\frac{1}%
{\sqrt{5}}$, $\frac{1}{\sqrt{10}}$, $\frac{1}{\sqrt{22}}$ and the red dashed
lines from top to bottom correspond to $\left\vert \alpha\right\vert =\frac
{2}{\sqrt{5}}$, $\frac{3}{\sqrt{10}}$, $\sqrt{\frac{21}{22}}$.}%
\label{Figure4}%
\end{figure}The figure shows that in the range of larger acceleration, the
loss of $\pi$-tangle depends only on the initial value of the degree of
entanglement. This shows that the response of $\pi$-tangle to $r$ is different
from logarithmic negativity for bipartite state because the latter does depend
on the choice of values of $\alpha$ and $\sqrt{1-\alpha^{2}}$. However, for
smaller values of acceleration, it does degrades, like the logarithmic
negativity for bipartite states, along two different trajectories by
interchanging the values of $\alpha$ and $\sqrt{1-\alpha^{2}}$. For every
value of initial entanglement, it has a nonvanishing value at infinite
acceleration. The notable feature of figure ($4$) is that, unlike bipartite
entanglement, the tripartite entanglement for scalar field degrades slowly
with acceleration than for Dirac field and it always remains finite in the
limit of larger values of $r$.

\section{Summary}

In this paper, we have investigated the entanglement behavior of nonmaximal
tripartite quantum states in both fermionic and bosonic systems when one of
the parties is traveling with a uniform acceleration. Rindler coordinates are
used for the accelerating party. The behavior of entanglement against the
acceleration parameter and the initial entanglement parameter is quantified
using $\pi$-tangle.

It is shown that the entanglement in tripartite GHZ states does not only
depend on the acceleration and initial entanglement in the states but also
depends, for the same initial entanglement, on the probability amplitudes of
the bases vectors. The one-tangles corresponding to accelerated observer, in
both bosonic and fermionic cases, strongly depends on the entanglement
parameter $\alpha$. However, in the fermionic case, it never vanishes for any
values of $\alpha$ even in the limit of infinite acceleration. Whereas in
bosonic case, regardless of the value of $\alpha$, it vanishes in the range of
infinite acceleration. The two-tangles, in both cases, are always zero, which
means that the acceleration and the degree of initial entanglement do not
affect the entanglement behavior of any of the sub-bipartite systems.

The response of $\pi$-tangle to $r$ and $\alpha$ in the two cases is
considerably different. In fermionic case, for the same initial entanglement,
it strongly depends on the values of $\alpha$ and $\sqrt{1-\alpha^{2}}$. The
difference in degradation against $r$, by interchanging the values of
probability amplitudes, increases with increasing acceleration. However, some
fraction of $\pi$-tangle always survives for all values of $\alpha$ even in
the limit of infinite acceleration. For bosonic case, in the range of large
values of $r$, the $\pi$-tangle just depends on the of initial entanglement,.
However, for small values of $r$, its degradation is different by
interchanging the values of probability amplitudes. Amazingly unlike bipartite
entanglement, the $\pi$-tangle in fermionic case degrades quickly against the
acceleration as compared to bosonic case. The survival of tripartite
entanglement may be used to perform different quantum information task in
situations where execution of such task through bipartite entanglement fails,
for example, between inside and outside of the black hole.

\end{document}